\documentclass[12pt]{article}
\usepackage{amsmath}
\usepackage{bm}
\usepackage{graphicx}
\usepackage{authblk}
\usepackage[a4paper]{geometry}

\begin{document}

\title{Field-induced upward bending of a magnetoelastomer cantilever residing on \newline a horizontal plane:
        an unconventional cilium}

\author[1]{O. V. Stolbov}
\author[2]{G. V. Stepanov}
\author[1]{Yu. L. Raikher}%
\affil[1]{Institute of Continuous Media Mechanics, Russian Academy of Sciences, Ural Branch, Perm 614018, Russia}
\affil[2]{State Research Institute for Chemical Technologies of Organoelement Compounds, Moscow 111123, Russia}

\maketitle

\begin{abstract}
      The mechanical response of a cantilever made of a magnetoactive elastomer (MAE) that is positioned on a solid plane surface and is subjected to a uniform magnetic field is studied.
      The MAE is of the magnetically soft type, i.e., its filler particles become magnetized only in the presence of external field.
      Test observations with the applied field normal to the plane reveal two possible perturbed configurations of the cantilever: either its free end just bends upward or the cantilever folds into an arc, so that its free end does not detach from the supporting plane.
      The tests evidence that under cyclic variation of the field both deformation modes exhibit quite a wide bistability region, i.e., a magnetomechanical hysteresis takes place.
      Theoretical analysis shows that the cantilever bending scenario is indeed similar to that of the first-order transition.
      However, unlike the customary case, hereby the initial state never ceases to exist and remains stable against infinitesimal perturbations under arbitrary strong fields.
      Although this conclusion is valid only for an ideal situation, its nontrivial consequence is that the field value under which the transition to the bent state occurs is, in fact, unpredictable.
      The discovered effect is essentially different from the previously investigated behaviour of MAE cantilevers under an in-plane field; there the transition is of the second-order kind and is completely reversible.
      Finally, if to consider a horizontal MAE cantilever as an element of an assembly of magnetic cilia, it turns out that it behaves in the way opposite to that of conventional systems where in the initial state the MAE cilia stand normally to the supporting surface.
\end{abstract}

\vspace{2pc}



\section{Introduction} \label{sec:1}
     The ability of magnetoactive elastomers (MAEs) to undergo significant deformations in relatively low fields has brought one- and two-dimensional (1D and 2D) elements---filaments, rods, thin-walled tubes, plates, membranes---into the focus of applicational interest.
     These anisometric objects are particularly easy to deform in the direction perpendicular to their main dimensions.
     This ability ensures a bright future for MAE elements as field-controlled grippers \cite{GaWa_ACS-AMI_16,HeBu_AMT_24}, actuators and dispensers \cite{BeGo_SMS_16,JaAw_SMS_18}, cantilevers with tuned frequency \cite{BeRa_SMS_17}, various sensors and resonators \cite{LoLo_SMS_11,MiHe_SMS_19,HuZh_IJSS_22}; whereas, if both ends of the MAE beam are free, one may get remotely controlled crawlers \cite{ChCa_AMT_20} and, more broadly, diverse types of soft microrobots \cite{KiPa_SR_19,WuHu_MM_20,KiZh_ChemR_22,ZhHa_PNAS_24}.
     Collectives of miniature rod-like MAE elements are worth special noting as they may be arranged in the carpets of magnetically-controlled artificial cilia and brushes \cite{AlKo_AdvMater_15,BaGa_JMMM_23}. 

     Low-dimensional MAEs demonstrate a high deformation response not only in a gradient field, where the sample evidently experiences the action of a volumetric force, but also under a uniform field, where the average force is absent.
     Examples include a thin membrane clamped at its rim in a perpendicular field \cite{RaSt_JP-D_08} or a cantilever positioned on a horizontal plane in a horizontally oriented normal field \cite{KaDz_SMS_21,KaDz_SMS_23}.

     For magnetically soft MAEs, in which the filler particles do not possess the magnetic moments of their own, the initial magnetization and the initial deformation response to a uniform field $\bm{H}$ are always quadratic in the magnitude of the latter.
     Therefore, the deformation of samples with symmetry axes under the action of a field perpendicular to one of the axes develops according to a scenario of transition from a high-symmetry to a low-symmetry state and is degenerate with respect to field inversion.
     In this case, a convenient method for theoretical description is the Landau expansion formalism, which allows the emerging effect to be interpreted as a second-order phase transition, for which a suitable characteristic is chosen as the order parameter.
     When studying the magnetomechanics of MAE samples, a natural candidate for this role is the displacement magnitude that arises as a result of the field action.
     In the case of a membrane, this is the deflection of the resulting 'dome' \cite{RaSt_JP-D_08}, and in the case of a cantilever dwelling on a plane, it is the deviation of its free end from the initial straight configuration \cite{KaDz_SMS_21,KaDz_SMS_23}.
     Indeed, these parameters have zero values for unmagnetized samples and grow up only under the action of a uniform field $\bm{H}_0$ thereby reducing the geometric symmetry of the sample.

\begin{figure}[ht]
     \centering
     \includegraphics[width=0.85\textwidth]{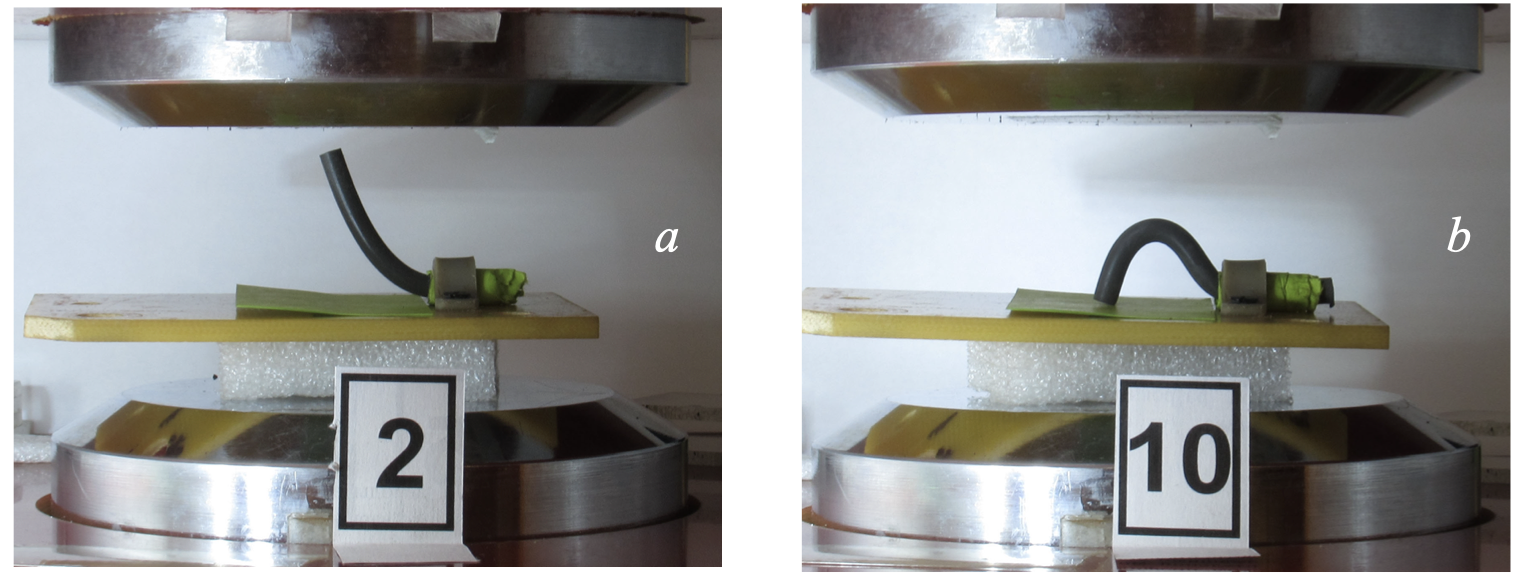}
     \caption{Experimentally observed deformation modes of the MAE rods residing on a horizontal plane in a magnetic field directed vertically.
     The lengths are 36\,mm, Young moduli are about 1\,MPa, carbonyl iron content: 40\,wt.\%{} (\emph{a}) 60\,wt.\%{} (\emph{b}).
     \label{fig:fig_01}}
\end{figure}

     In the present work, we consider the case of a magnetically soft MAE cantilever (a rod or beam) lying on a horizontal plane under the action of a magnetic field directed vertically.
     The general, insofar just illustrative, evidence is given in Fig.\ \ref{fig:fig_01}, which renders two alternative deformation patterns.
     The qualitative analysis of the experiment observations enables one to establish three main features of the effect.
     First, it is of a threshold type: on attaining  a certain field value, the inflection at once acquires a considerable magnitude that changes but weakly upon further growth of the field. 
     Second, on diminution of the field strength below the onset one, the inflection decreases gradually and drops down only after a substantial reduction of the field.
     Third, the inflection develops along one of the two possible ways: either it is a simple bend or an arc (hairpin) configuration.
     
     Therefore, it turns out that the situation with the vertical (upward) deformation of the cantilever is qualitatively different compared to its deformation under a horizontal (in-plane) field.
     Whereas in the in-plane (independent of gravity) case, the deformation evolves according to a second-order transition scenario, which does not imply bistability, in the case of vertical field, the observed deformational transition exhibits significant hysteresis.
     The explanation follows from two factors which make the upward and downward directions non-equivalent: first, the presence of gravity force, and, second, the presence of an impermeable barrier.
     As our analysis shows, in terms of the magnitude of the applied field, the hysteresis (bistability) region may, in principle, have an infinite width.
     The solution presented in this work is obtained by the variational method and, therefore, is not mathematically exact.
     However, the results presented below help to explain the origin of each of the characteristic configurations shown in Fig.\ \ref{fig:fig_01}.

\section{MAE-cantilever on a horizontal plane \label{sec:02}}
\begin{figure}[ht]
     \centering
     \includegraphics[width=0.96\textwidth]{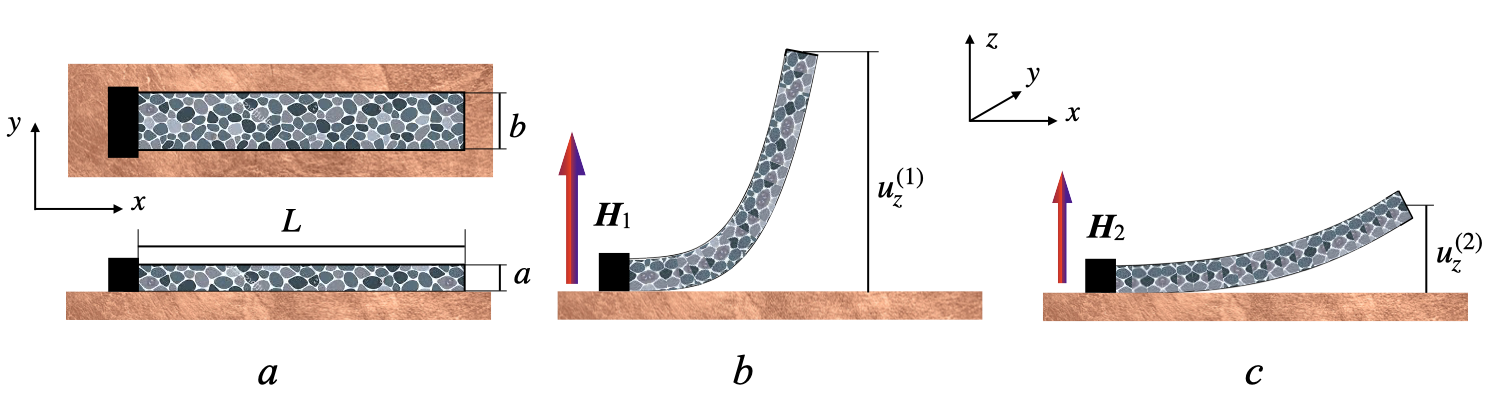}
     \caption{Schematic representation of deformational response of the magnetically soft MAE cantilever.
     \label{fig:fig_02}}
\end{figure}

     The initial state of a magnetically soft MAE cantilever and the sequence of its configurations under simple bending are shown schematically in Fig.\ \ref{fig:fig_02} for the case shown in Fig.\ \ref{fig:fig_01}{\it a}.
     Under external field $\bm{H}_0$ directed perpendicular to the rod axis, it retains its initial configuration (Fig. \ref{fig:fig_02}{\it a}) up to a certain value $H_{0\ast}$.
     Then, the free end abruptly inflects upward (Fig. \ref{fig:fig_02}{\it b}) and after that but slightly changes its configuration as the field strength increases.
     Upon decreasing the field, the deformation reduces gradually (Fig.\ \ref{fig:fig_02}{\it b} $\Rightarrow$ Fig.\ \ref{fig:fig_02}{\it c} $\Rightarrow$ Fig.\ \ref{fig:fig_02}{\it a}).
     Full recovery of the horizontal configuration takes place after the field is reduced to a certain value $H_{0c}<H_{0\ast}$.
     Thus, significant deformational hysteresis occurs.

     For theoretical analysis of the problem, the MAE sample is modelled as a cantilever, i.e., a lengthy object clamped at one its end; it may have circular or rectangular cross-section.
     In our scheme, the clamp outer edge is the end of the sample as well. 
     The MAE itself is characterised by density $\rho$, Young modulus $E$, Poisson ratio $\nu$ and initial magnetic susceptibility $\chi$.
     It is assumed that the distributions of all these material parameters within the sample are homogeneous and insensitive to external influences.
     The rod has length $L$, cross-section area $S$, and resides on a solid horizontal plane ($z=0$).

\section{Plain bending. Qualitative analysis of the model} \label{sec:03}
\subsection{Energy of a MAE rod under a magnetic field}  \label{sec:03.1}
     A MAE rod deforms due to the change of its energy under the action of applied field.
     To obtain the corresponding contribution, let us consider an element $\Delta L$ of the cantilever length and assume that this element is a fragment of an elongated ellipsoid of revolution, whose major axis afterwards would be extended to infinity.
     In this case—see Appendix—the volume density of magnetic energy may be presented as
\begin{equation} \label{eq:01}
     W_{\rm magn}=Q\!\cdot\!\left(\bm{\tau}\cdot\bm{H}_0\right)^2,
\end{equation}
where $\bm{H}_0$ is the applied field, $\bm{\tau}$ is the unit vector in the direction of major axis of the ellipsoid, whereas the shape factor depends only on the MAE magnetic susceptibility and in the limit of a long ellipsoid is given by
 \begin{equation} \label{eq:02}
     Q=\pi\chi^2/(1+2\pi\chi).
 \end{equation}
     
     The total energy of the cantilever with cross-section area $S$ is presented as the sum of three contributions: elastic, magnetic, and gravitational ones:
\begin{equation} \label{eq:03}
     U={\textstyle \frac12}EJ\int_0^L \left(\frac{d^2u_z}{dx^2}\right)^{\!\!2} dx-Q\int_V(\bm{\tau}\!\cdot\!\bm{H}_0)^2 dV+\rho g\int_V u_z dV;  
\end{equation}
the coordinate frame used is shown in Fig. \ref{fig:fig_02}.
     In formula (\ref{eq:03}) $J$ is the 'geometric' moment of inertia per unit length of the rod, $V=SL$ its volume, $g$ the acceleration of gravity, and $u_z$ is the displacement of the rod free end along vertical axis $Oz$.

     Upon Introducing dimensionless variables
\begin{equation} \label{eq:04}
     \tilde{x}=x/L, \quad \tilde{u}_z=u_z/L, \quad \tilde{H}_0 = H_0\sqrt{Q/E},
\end{equation}
one may present the energy integral (\ref{eq:03}) in the form
\begin{equation} \label{eq:05}
     \tilde{U}=U/EV=\int_0^1d{\tilde x}\left[{\textstyle \frac12}\beta\left(\frac{d^2{\tilde u}_z}{d{\tilde x}^{2}}\right)^{\!\!2}\!\!-(\bm{\tau}\!\cdot\!{\tilde{\bm{H}}}_0)^2+\alpha{\tilde u}_z\right],
\end{equation}
with dimensionless parameters $\beta=J/(L^2S)$ and $\alpha=\rho gL/E$.

     Denoting the angle between vectors $\bm{\tau}$ and $\bm{H}_0$ by $\theta$, so that ${\cot}\,\theta=d{\tilde u_z}/d{\tilde x}$, one finds
\begin{displaymath}
     (\bm{\tau}\!\cdot\!{\tilde{\bm{H}}}_0)^2/{\tilde H}_0^2\!=\!\cos^2\theta\!=\!\frac {\cot^2\theta}{1+\cot^2\theta}\!=\!\left(\frac{d{\tilde u}_z}{d{\tilde x}}\right)^{\!\!2}\!\!\Big/\left[1\!+\!\left(\frac{d{\tilde u}_z}{d{\tilde x}}\right)^{\!\!2}\right]\!\approx\!\left(\frac{d{\tilde u}_z}{d{\tilde x}}\right)^{\!\!2}\!\!-\left(\frac{d{\tilde u}_z}{d{\tilde x}}\right)^{\!\!4},
  \end{displaymath}
that yields the final expression
\begin{equation} \label{eq:06}
     \tilde{U}=\int_0^1d{\tilde x}\left\{{\textstyle \frac12}\beta\left(\frac{d^{2}{\tilde u}_z}{d{\tilde x}^{2}}\right)^{\!2}\!\!-H_0^2\left[\left(\frac{d{\tilde u}_z}{d{\tilde x}}\right)^{\!\!2}\!\!-\left(\frac{d{\tilde u}_z}{d{\tilde x}}\right)^{\!\!4}\right]+\alpha{\tilde u}_z\right\}.
\end{equation}

     It is worthwhile to ascertain the values of dimensionless parameters that are characteristic of experimental samples like those shown in Fig.\ \ref{fig:fig_01}.
     For a rod with circular cross-section of radius $a$, one has $\beta_{\rm cyl}={\textstyle \frac14}(a/L)^2$; for a rod with square cross-section of size $2a$, it yields $\beta_{\rm beam}=\frac1{3}(a/L)^2$.
     Setting $a=2.25$\,mm, $L=36$\,mm, $\rho\approx3$\,g/cm$^3$, $E\sim10^7$\,dyn/cm$^2$, and $g=981$\,cm/s$^2$, one obtains the estimate $\alpha\approx1.1\times10^{-3}$.
     For the shape factor $Q$, formula (\ref{eq:01}) with $\chi=0.2$ gives $Q\approx0.05$.

\subsection{Cantilever in the absence of spatial restrictions} \label{sec:03.2}
\begin{figure}[ht]
\centering
\includegraphics[width=0.45\textwidth]{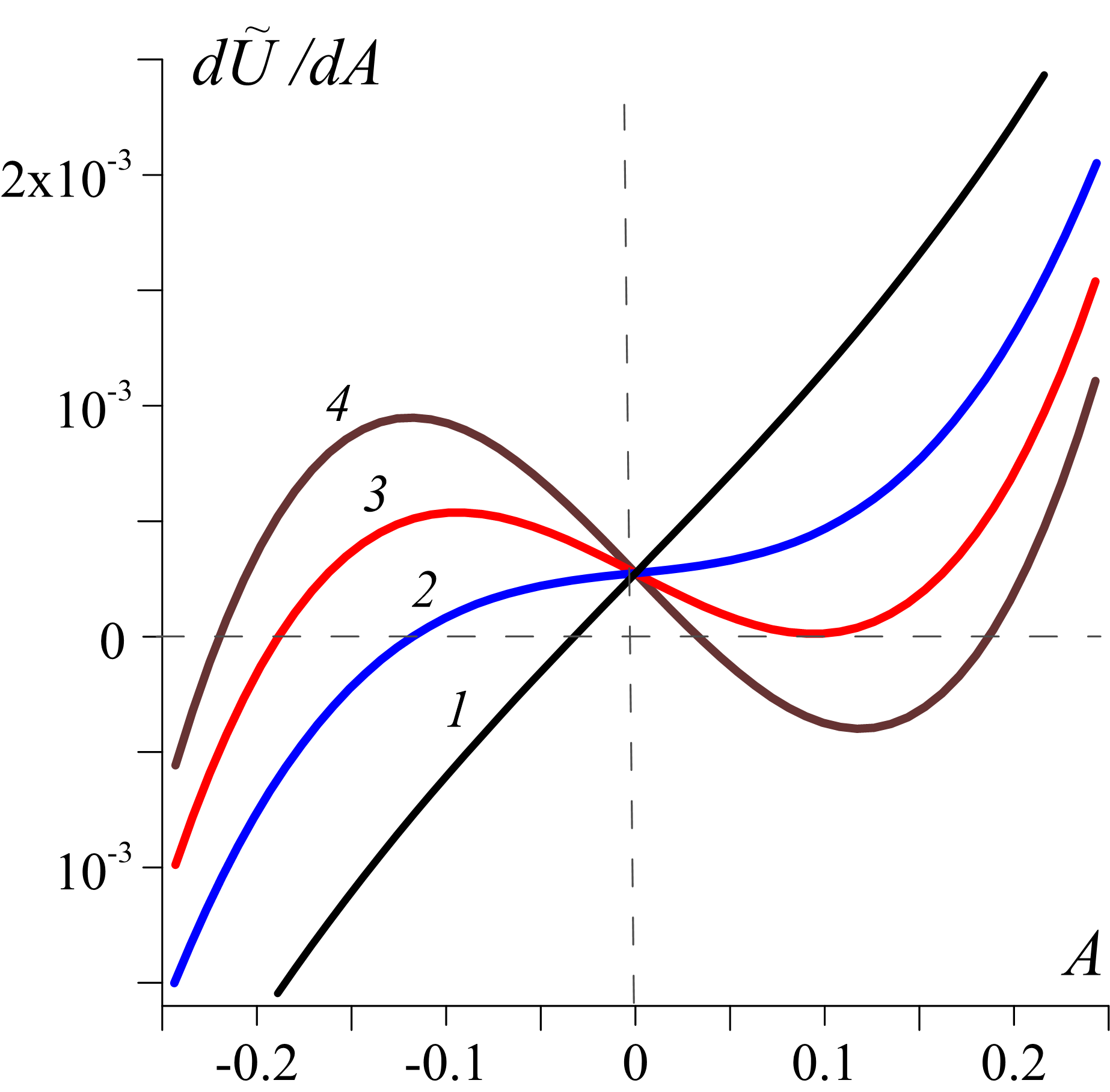}
\caption{Derivative $d{\tilde U}/dA$ for the same set of parameters.
            \label{fig:fig_04}}
\end{figure}

     Consider simple bending of a cantilever whose left end ($x=0$) is fixed; in this section we do not impose any spatial constraints.
     As the most simple approximation of the emerging configuration, function $\tilde{u}_z=A\tilde{x}^3$, is taken, which qualitatively reproduces a simple bend observed in the experiment.
     Substituting this profile into (\ref{eq:06}) and integrating over the rod length, one gets
\begin{equation} \label{eq:07}
\tilde{U}=9\tilde{H}_0^{2} A^4+\left(6\beta-{\textstyle \frac95}\tilde{H}_0^2\right)A^2+{\textstyle \frac14}\alpha A.
\end{equation}
     The field dependences of this function and its derivative are plotted in Figs.\ \ref{fig:fig_04} and \ref{fig:fig_05} for the set of material parameters referred to in section \ref{sec:02}.
     Note that at this stage of analysis, both upward and downward bends of the cantilever are allowed.

\begin{figure}[ht]
\centering
\includegraphics[width=0.45\textwidth]{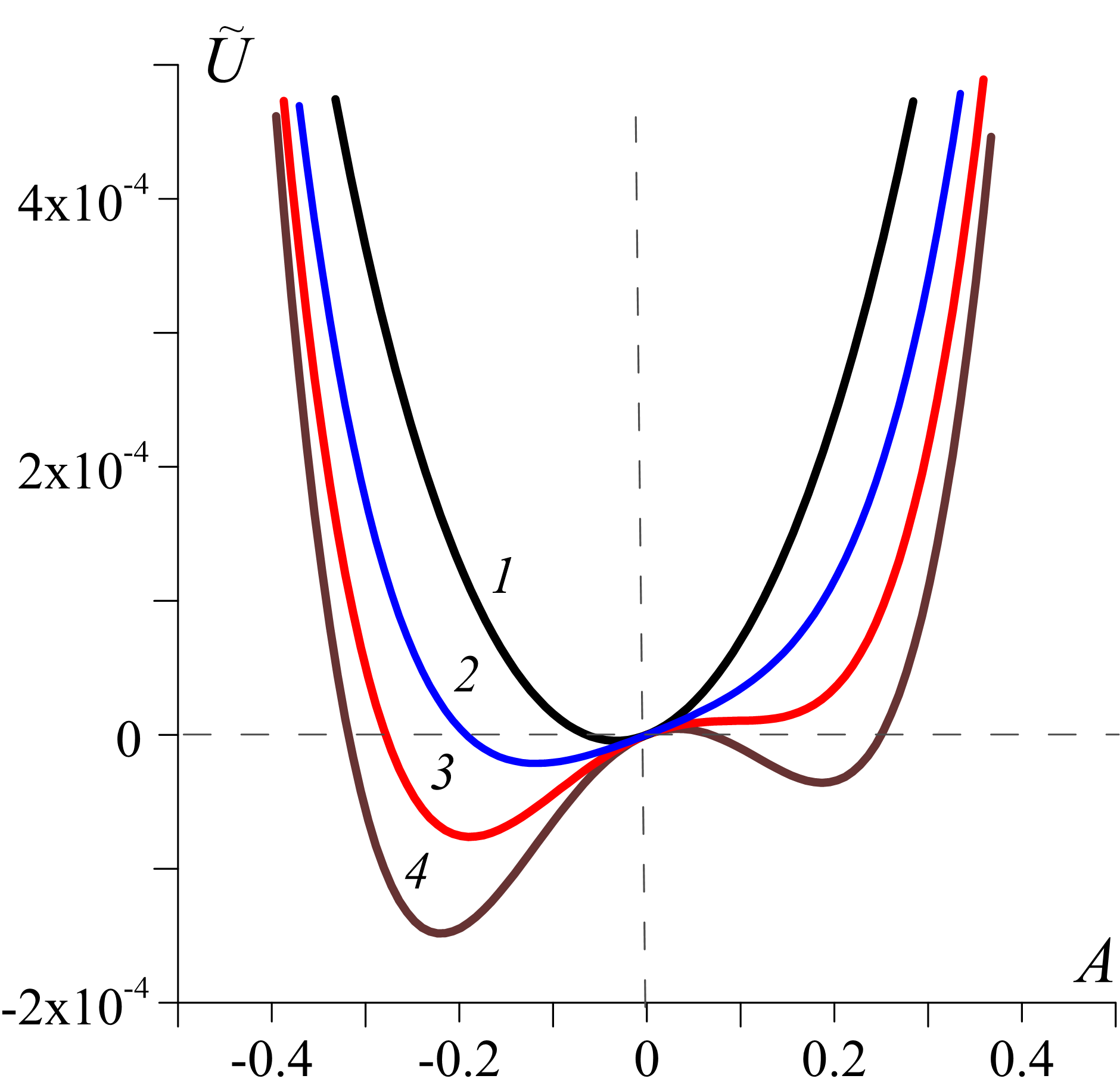}
\caption{Energy function (\ref{eq:07}) for parameter values $\alpha=1.1\times10^{-3}$ for fields $\tilde{H}_0=0.03$ 
             (\emph{1}), $0.055$ (\emph{2}), $0.0665$ (\emph{3}), and $0.075$ (\emph{4}). 
             \label{fig:fig_03}}
\end{figure}

     Under the field increase, the energy curves in Fig.\ \ref{fig:fig_03} change in a comprehensible way.
     In relatively weak fields, ${\tilde U}$ has a single minimum located at negative values of amplitude $A$: the cantilever sags under its own weight, see curve \emph{1}.
     Under the field increase, the influence of magnetic contribution in $\tilde U$ grows compelling the cantilever to a higher extent align with the direction of applied field.
     Due to that, the downward deflection augments: the corresponding energy minimum shifts towards greater negative values of $A$ and becomes deeper, see curve \emph{2}.
     A further field increase not only deepens the minimum at $A<0$ but also brings in another minimum for which $A>0$, see curve \emph{3} in Fig.\ \ref{fig:fig_03}.
     This takes place at some value ${\tilde H}_0={\tilde H}_{0c}$ when the derivative $d{\tilde U}/dA$ first reaches zero, see curve \emph{3} in Fig.\ \ref{fig:fig_04}.
     Under yet stronger fields, as shown by curve \emph{4} in Fig.\ \ref{fig:fig_03}, the cantilever becomes bistable, since for it now two different equilibrium states become available.
     The first one corresponds to a deep minimum at $A<0$ and, thus, means further sagging.
     The alternative state is associated with the energy minimum at $A>0$ and means an upward bend.
     It is important to note, however, that a free cantilever cannot switch into this state spontaneously; to attain it, a certain potential barrier must be overcome.
     Meanwhile, the MAE rod is a macroscopic system where the effect of thermal fluctuations is virtually absent.
     This implies that, the cantilever would never attain the 'positive' potential well---and, thus, bend upward---unless some external finite-amplitude perturbation would be exerted: a force impact or forced deformation, for example.

\subsection{Cantilever residing on a plane} \label{sec:03.3}
\begin{figure}[ht]
 \centering
 \includegraphics[width=0.42\textwidth]{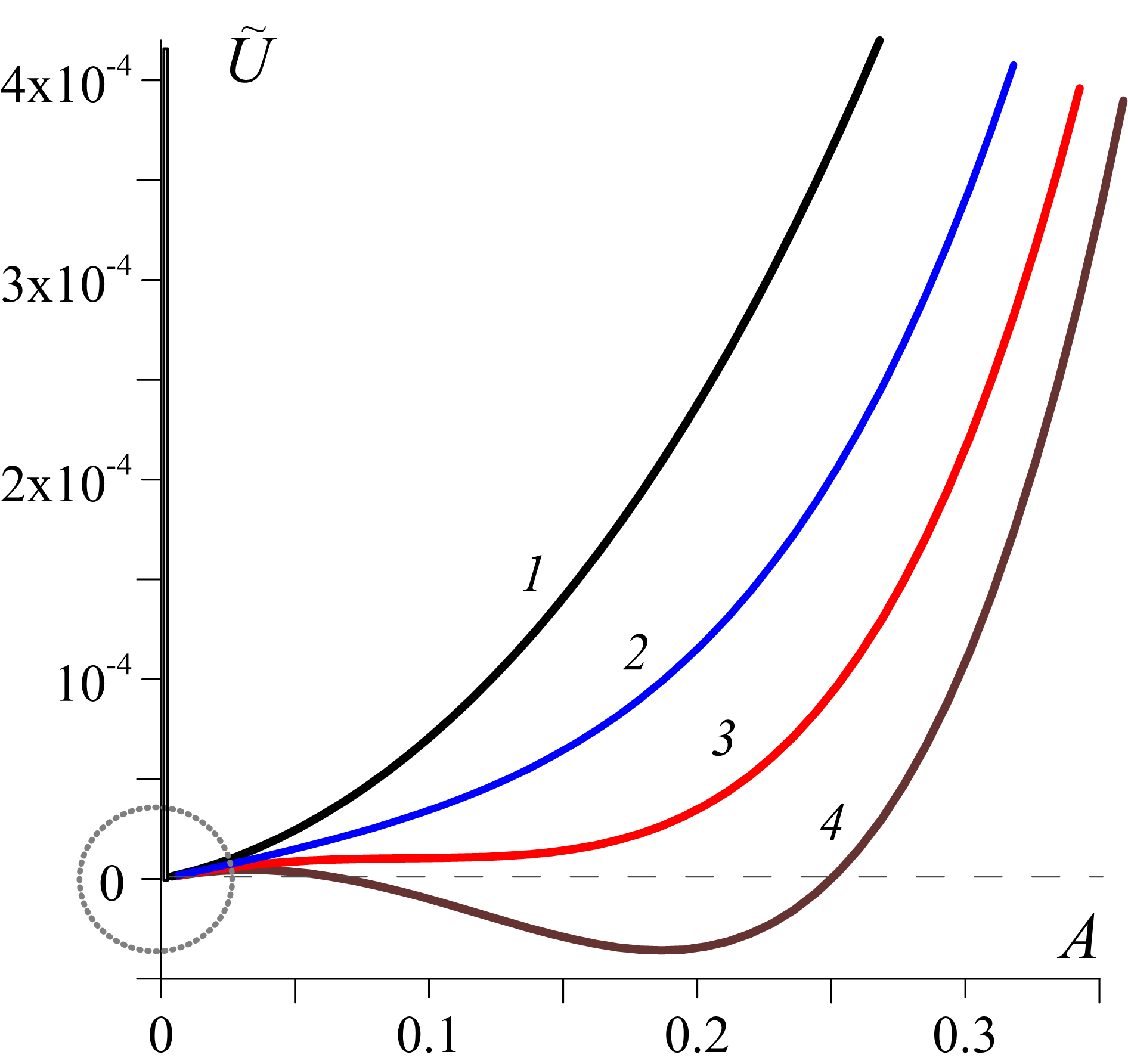}
 \caption{Energy of a model cantilever residing on a horizontal plane as a function of the magnetic field strength; all parameter values are the same as in Fig.\
               \ref{fig:fig_04}; the vertical line $A=0$ represents the left wall of the potential well (solid plane); the dashed circle marks the location of the ever present
               energy 'pocket'.
               \label{fig:fig_05}}
\end{figure}

     We modify the above-considered problem statement by imposing a constraint that follows from the fact that the MAE rod resides on an impermeable solid plane, i.e., downward bend is impossible: $A\geq0$.
     Therefore, in the plots of Fig.\ \ref{fig:fig_03} only the region with non-negative amplitude remains accessible; formally, this implies that the value ${\tilde U}(A=0-\varepsilon)$ with $\varepsilon$ being an infinitely small quantity, should be considered as ${\tilde U}=\infty$.
     Given these conditions, a new potential well---the minimum located at the left boundary of the now-imposed domain of $A$---appears on the energy curves in Fig.\ \ref{fig:fig_04}, as shown in Fig.\ \ref{fig:fig_05}.
     If the field is below a certain critical value ${\tilde H}_{0c}$, the energyscape comprises only this single minimum ('pocket') located at $A=0$; it corresponds to the initial straight horizontal configuration.
     Only after the field grows up to ${\tilde H}_0\geq{\tilde H}_{0c}$, there emerges, exactly as in section \ref{sec:03.2}, another minimum that corresponds to the upward bend of the cantilever.

     To find the stationary configurations, one should set to zero the derivative $d\tilde{U}/dA$ and solve the resulting cubic equation
\begin{equation} \label{eq:08}
    36{\tilde H}_0^{2}A^3+6\left(2\beta-{\textstyle \frac35}{\tilde H}_0^2\right)A+{\textstyle \frac14}\alpha=0.
\end{equation}
     In principle, function $A({\tilde H}_0)$ could be found from (\ref{eq:08}) analytically, however, this way is rather cumbersome.
     However, if to treat equation (\ref{eq:08}) as a definition of the inverse function:
\begin{equation}  \label{eq:09}
 \tilde{H}_0 (A) = \frac16\sqrt{\frac{5(\alpha+48\beta A)}{2A\left(1-10A^{2}\right)}},
\end{equation}
the 'equation of state' $A({\tilde H}_0)$ is obtained in a far easier way.

\begin{figure}[ht]
 \centering
 \includegraphics[width=0.38\textwidth]{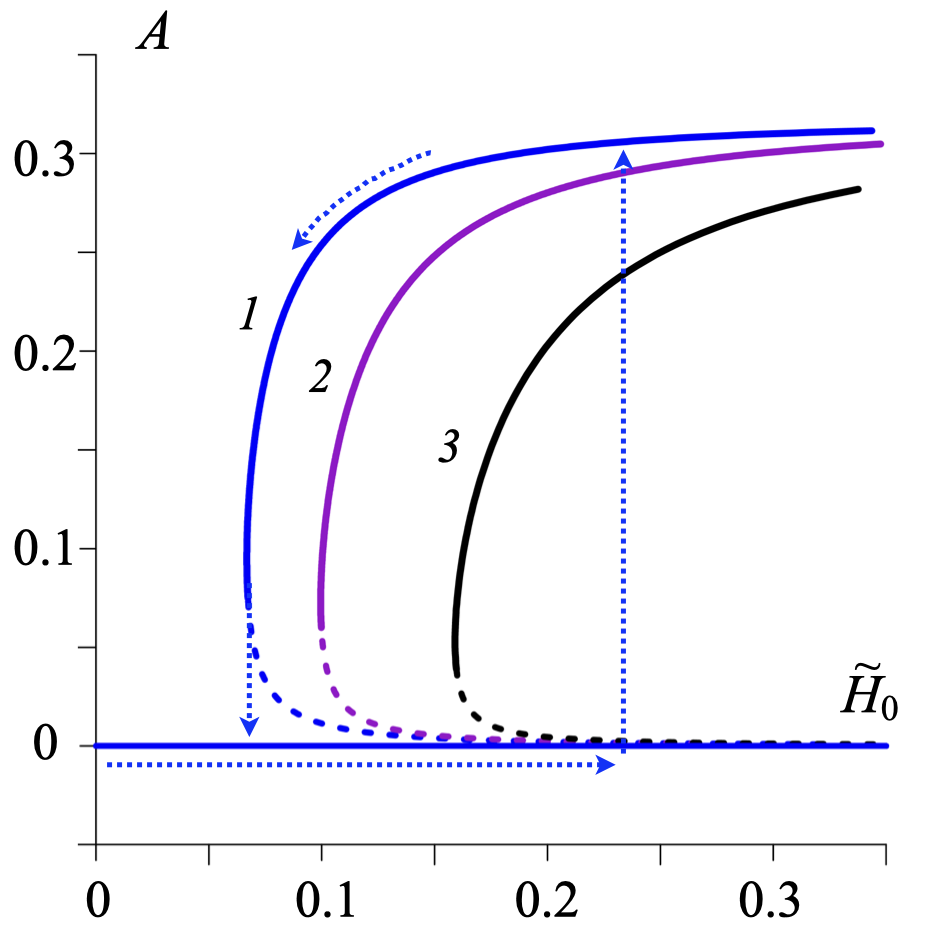}
 \caption{'Equation of state' for the model cantilever with respect to bending deformation for various aspect ratios: $L/a=16$ (\emph{1}), 10 (\emph{2}), and 6
              (\emph{3}); the values of other parameters are the same as those indicated in section \ref{sec:02}; dashed lines correspond to unstable branches.
              \label{fig:fig_06}}
\end{figure}

     The plots of function $A({\tilde H}_0)$ are shown in Fig.\ \ref{fig:fig_06}, solid lines present stable (stationary) states, dashed lines show unstable ones.
     Note that Fig.\ \ref{fig:fig_06} supports the expected trend: the shorter the cantilever, the higher the field required for its transition to a bent state.

     At a first glance, the set of lines in Fig.\ \ref{fig:fig_06} resembles typical $S$-shaped curves characteristic of a first-order phase transition.
    However, there is a qualitative difference: here the horizontal line $A({\tilde H}_0)=0$ that corresponds to the initial straight configuration of the cantilever does not have any end point, it goes to infinity.
    In other words, the high-symmetry state survives, i.e., remains stable against infinitesimal perturbations, under magnetic field of arbitrary strength.
    Indeed, as can be extrapolated from the curves in Fig.\ \ref{fig:fig_04} or \ref{fig:fig_06}, the energy 'pocket' at $A=0$—see the dashed circle in Fig.\ \ref{fig:fig_05}—is ever present although, as the applied field increases, this minimum becomes more shallow.
    In other words, it disappears only at ${\tilde H}_0\rightarrow\infty$.
    As mentioned above, the considered system is insensitive to thermal fluctuations.
    This implies that the instability (bending) of the cantilever may be induced only by finite-amplitude perturbations.

     It is important to note, however, that this conclusion applies in full only to the model case, which assumes a perfectly homogeneous composition of the MAE, a geometrically flawless cantilever, an absolutely flat substrate, and a completely uniform magnetic field around the sample.
    In a real situation, any violation of one (or several) of these conditions would act as a static fluctuation, provoking breakup of the high-symmetry state.
    This, from our point of view, is the reason why in the experiments the cantilever bending occurs at some breakup field ${\tilde H}_{0\ast}$ as if spontaneously, i.e., in the absence of any intentional stimulations.
    The transition undergoes quite abruptly, the cantilever free end at once lifts to a significant height.

    Unlike the onset of bending that is, most probably, caused by a combination of weak and poorly accounted for factors, the return of the cantilever to its initial configuration under reduction of the applied field is gradual and quite predictable.
    Indeed, at this stage the energy excess in the system is substantial, so that small imperfections which had once provoked the bend are insignificant.
    Under a slow decrease of the field, the cantilever straightens smoothly.
    As a certain critical value ${\tilde H}_{0c}$ is attained, the cantilever recovers its initial straight horizontal configuration. 
    The on/off field cycle is outlined in Fig.\ \ref{fig:fig_06} with dotted arrows; the point of the instability onset under increasing field is chosen arbitrarily.

    The value ${\tilde H}_{0c}$ can be determined by finding the minimum of function $\tilde{H}_0 (A)$ for $A>0$.
    Since direct differentiation of formula (\ref{eq:09}) leads to a cumbersome nonlinear expression, it is more convenient to solve the equation $d\tilde{H}_0/dA=0$ numerically.
    The obtained dependencies of ${\tilde H}_{0c}$ on the MAE's elastic modulus and on the cantilever length have the anticipated shape form: for a given field strength, the longer the cantilever the greater its bending; the stiffer the MAE the harder it is to deform.

\section{\normalsize Simple bending of a cantilever. Three-parameter approximation} \label{sec:04}
     The main advantage of the simplified variational function used above is that it provides a very clear scheme for understanding how the MAE cantilever deforms under the increase/decrease cycle of the field. 
     However, this approximation is completely inadequate for quantitative description of the effect, primarily because the trial configuration contains only one degree of freedom---a single coefficient $A$. 
     To develop a more realistic approach, we turn to the classical problem of an elastic rod bending, whose exact solution for a non-magnetic cantilever is given by a fourth-order polynomial with three numerical coefficients \cite{LaLi_TE_86}. 
     Using this expression as a basis, we take the extended variational function in the form
\begin{equation}  \label{eq:10}
   {\tilde u}_z={\tilde x}^2(A{\tilde x}^2+B{\tilde x}+C),
\end{equation}
where all parameters are functions of ${\tilde H}_0$.
     Factor ${\tilde x}^2$ in formula (\ref{eq:10}) is necessary because at the fixed end of the cantilever (${\tilde x}=0$), conditions ${\tilde u}_z=0$ and $d{\tilde u}_z/d{\tilde x}=0$ must always hold.

     As for the cantilever only the half-space $z\geq0$ is allowed, that is, ${\tilde u}_z\geq0$, the coefficients in formula (\ref{eq:10}) should obey the following conditions: (\emph{i}) $C\geq0$ and (\emph{ii}) $A+B+C\geq0$. 
     One more restriction follows from the requirement of non-positivity of the discriminant of the quadratic trinomial in (\ref{eq:10}): (\emph{iii}) $B^2-4AC\leq0$. 
     Each of the conditions (\emph{i})--(\emph{iii}), if only equality is maintained, determine in the three-dimensional parameter space $(A,B,C)$ an open surface that separates the 'allowed' and 'forbidden' regions with respect to ${\tilde u}_z$.
    The geometric representations of the first and second conditions are planes, while the representation of the last condition is a second-order surface.
    The admissible representative points reside inside the space region determined by intersection of all these three 'allowed' half-spaces.
    The starting point that corresponds to the initial position of the cantilever at ${\tilde H}_0=0$, has coordinates $A=B=C=0$ and evidently belongs to all three boundary surfaces.
    
     Substituting the profile (\ref{eq:10}) into functional (\ref{eq:06}) and integrating over the rod length, one obtains energy ${\tilde U}(A,B,C)$ as a function of the variational parameters; analysis of its extrema allows one to find stable (minima) and unstable (maxima and saddle points) configurations of the cantilever.
     Due to the multidimensional nature of function ${\tilde U}(A,B,C)$, there is no convenient way to reproduce its energyscape graphically, similar to what is done in Figs.\ \ref{fig:fig_03} and \ref{fig:fig_05}.
     However, the existence of energy 'pocket' (fundamental stability of the initial horizontal position), see the circled area in Fig.\ \ref{fig:fig_05}, can be proven here as well.
     This follows from the fact that at the point $(0,0,0)$, the derivatives of $U$ with respect to all parameters are positive ($\sim\alpha$) and do not depend on ${\tilde H}_0$.
     In other words, the horizontal state of the cantilever---corresponding to the line $A=0$ in Fig.\ \ref{fig:fig_06}---survives under any applied field.
    
     Note that this conclusion is universal and does not depend on the number of degrees of freedom of the variational function if the latter is taken in polynomial form: ${\tilde u}_z=\Sigma_{k=0}D_k{\tilde x}^k$.
     Indeed, regardless of the polynomial order, the derivative of the rod energy with respect to the coefficient of any expansion term would necessarily be positive at the point $(0,0,0)$ whatever is ${\tilde H}_0$ strength.
     The physical reason is quite clear: as long as the middle line of the ideal MAE rod is not bent due to external influence, gravity will always keep the cantilever onto the plane and maintain its straight configuration.
     Of course, this conclusion does not apply to the case of an imperfect rod and/or incompletely uniform external conditions.
    
\begin{figure}[ht]
     \centering
\includegraphics[width=0.45\textwidth]{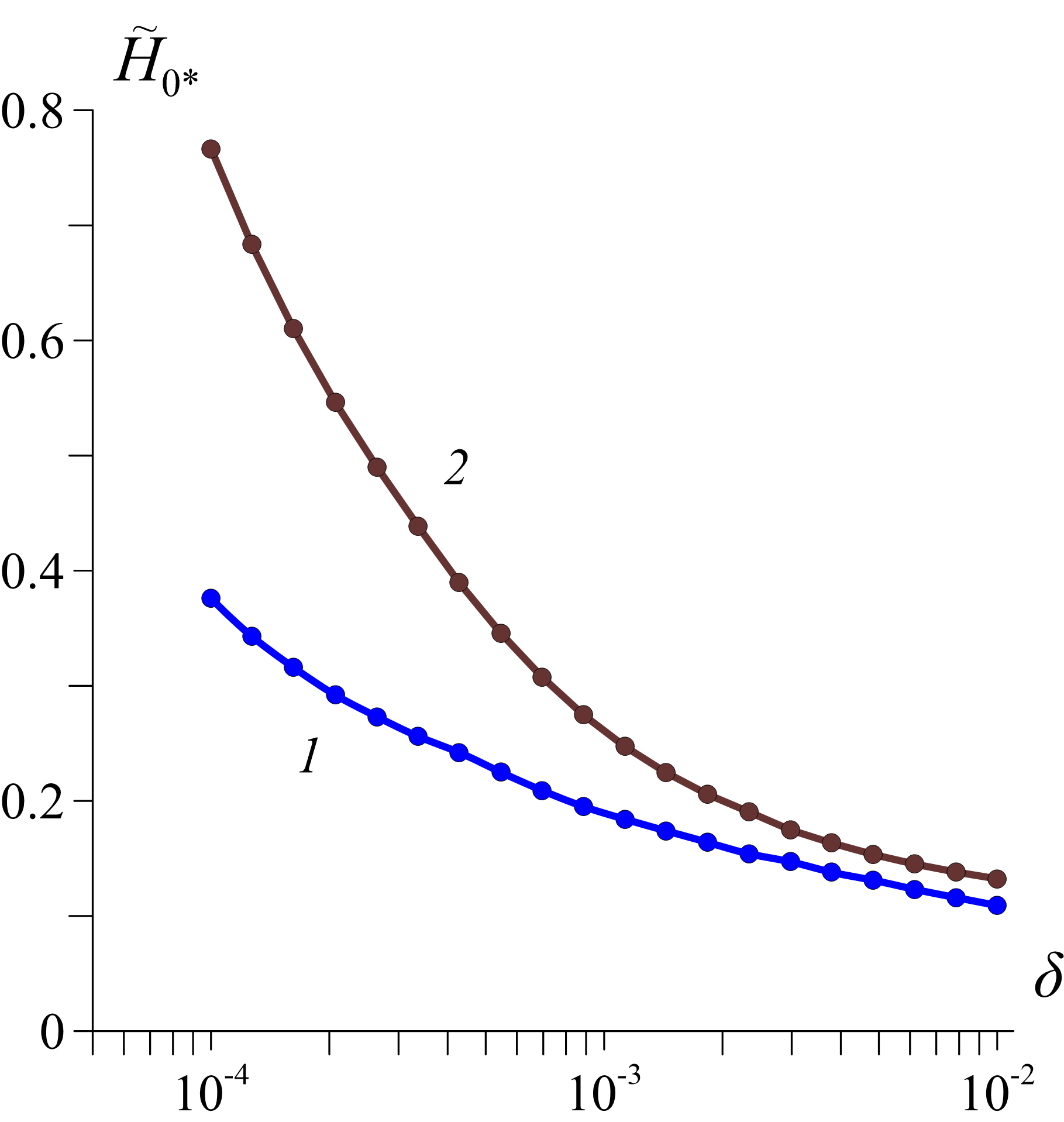}
    \caption{Dependence of the field ${\tilde H}_{0*}$ for the transition to a bent configuration on the magnitude of initial perturbation $\delta$ for simple (\emph{1}) and arc (\emph{2}) bendings.
    \label{fig:fig_07}}
\end{figure}

      Any imperfection---shape and geometry defects of the rod, non-planarity of the substrate, non-uniformity of spatial distributions of the magnetic filler and applied field, etc.---can become a source of finite-amplitude perturbation causing instability of the straight configuration of the cantilever.
     Unfortunately, within the framework of the present model, these contributions cannot be assessed.
     However, some idea of their influence can be obtained by considering the dependence of the breakup field ${\tilde H}_{0\ast}$ for a bent configuration on the magnitude of a controlled perturbation.
     As such, we introduce upward deviation of the free end of the rod: $\delta=(\Delta z/L)_{{\tilde x}=1}$.
     Curve \emph{1} in Fig.\ \ref{fig:fig_07} obtained in numerical calculations for the case of three-parameter variational function (\ref{eq:10}) shows the dependence ${\tilde H}_{0\ast}(\delta)$ thus describing 'provocation' of simple bending of the cantilever by weak lifting of its end.
     The accelerating growth of ${\tilde H}_{0\ast}$ with diminution of the perturbation magnitude supports the conclusion that straight configuration of a perfect rod is stable against infinitesimal perturbations under a field of any finite strength; note that the horizontal axis Fig.\ \ref{fig:fig_07} is logarithmic.
     
\section{\normalsize  Arc bending of a cantilever} \label{sec:05}
\subsection{\normalsize  Qualitative analysis} \label{sec:05.1}
     Approximation (\ref{eq:10}) is a fourth-order polynomial, which for a cantilever prone to arbitrary deformations should obey a single constraint: one end of the rod is fixed.
     In section \ref{sec:04}, to study simple bending, some additional conditions were imposed on the coefficients of formula (\ref{eq:10}).
     Meanwhile, in general, formula (\ref{eq:10}) is capable of describing multiple perturbed configurations of the cantilever.
     In particular, those where the maximum inflection develops in some internal point of the rod.

     Fig.\ \ref{fig:fig_01}\emph{b} shows such a situation: the cantilever is bent in an arc, so that its free end does not detach from the substrate.
     This configuration implies another geometric constraint: ${\tilde u}_z(1)=0$.
     Along the already developed line of qualitative analysis, to understand the origin of this mode it is convenient to take the variational function in the following single-parameter form
\begin{equation} \label{eq:11}
   \tilde{u}_z=KA(1-{\tilde x}){\tilde x}^3, \qquad K={\textstyle \frac{128}{27}}.
\end{equation}
     The numeric coefficient in this formula is introduced to make the maximum value of ${\tilde u}_z$ within the segment ${\tilde x}\in[0,1]$ equal to $A/2$, i.e., half of that for simple bending.

     The expression for the dimensionless energy and the equation determining the value of ${\tilde H}_{0c}$, which result from function (\ref{eq:11}), are
\begin{equation} \label{eq:12}
   {\tilde U}={\textstyle \frac{1}{5}}K\left[{\textstyle \frac{27}{143}}{\tilde H}_0^{2}K^3\!A^4+(12\beta-{\textstyle \frac37}{\tilde H}_0^2 )K\!A^2 +{\textstyle \frac14}\alpha A\right],
\end{equation}   
and     
\begin{equation} \label{eq:13}
   \frac{d\tilde{U}}{dA}={\textstyle \frac{1}{5}}K\left[{\textstyle \frac{108}{143}}{\tilde H}_0^{2}K^3A^3+2(12\beta-{\textstyle \frac37}{\tilde H}_0^2 )KA+{\textstyle \frac14}\alpha\right]=0,
\end{equation}   
so that the rearrangement of equation (\ref{eq:13}) gives
\begin{equation}  \label{eq:14}
   \tilde{H}_0 (A) = \frac12\sqrt{\frac{\alpha+96\beta KA}{KA\left(\frac67-\frac{108}{143}K^2\!A^{2}\right)}}.
\end{equation}

\begin{figure}[ht]
   \centering
   \includegraphics[width=0.46\textwidth]{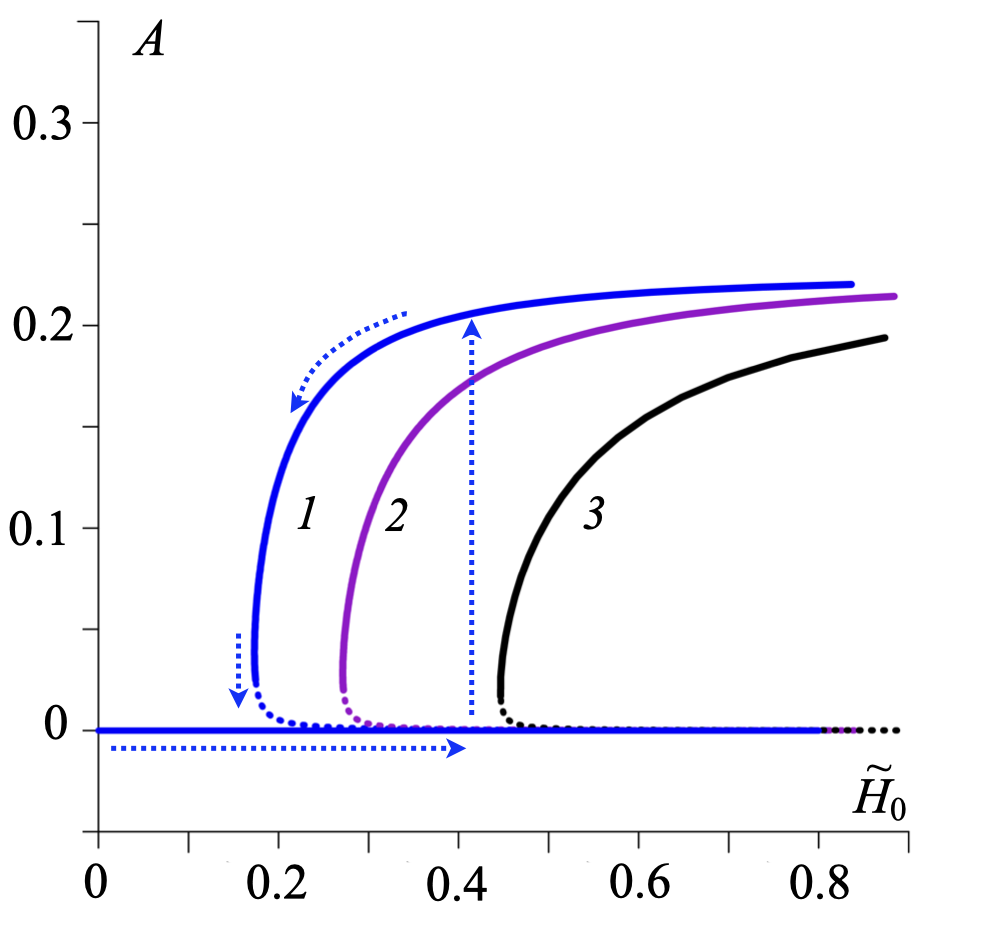} 
   \caption{`Equation of state' for the model cantilever with respect to arc deformation at different values of the dimensionless length $L/a$: 16 (\emph{1}), 10 (\emph{2}), and 6 (\emph{3}); values of other parameters are the same as those specified in section \ref{sec:02}; dashed lines indicate unstable branches.
   \label{fig:fig_08}}
\end{figure}

\begin{figure}[ht]
   \centering
   \includegraphics[width=0.46\textwidth]{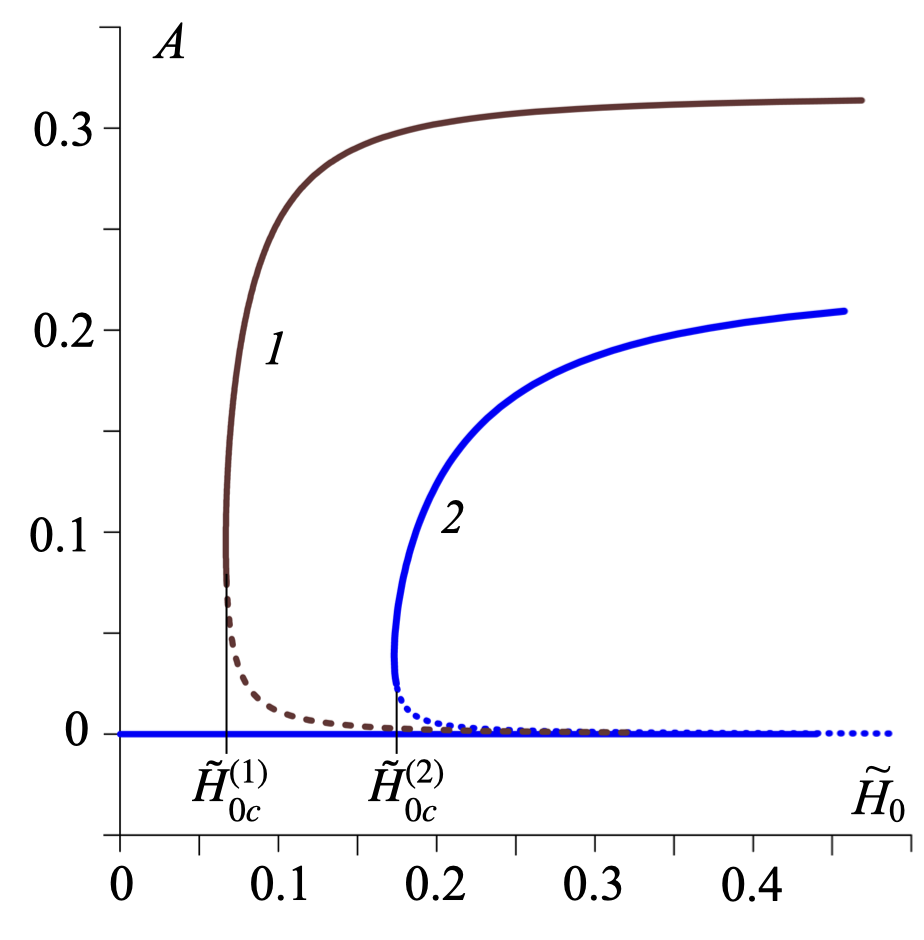} 
   \caption{Comparison of `equations of state' for the model cantilever with respect to bend (\emph{1}) and arc (\emph{2}) deformations at $L/a=16$; values of other parameters are the same as those specified in section \ref{sec:02}. 
   \label{fig:fig_09}}
\end{figure}

     The results of calculations similar to those performed in section \ref{sec:03} are presented in Fig.\ \ref{fig:fig_08}, whereas Fig.\ \ref{fig:fig_09} compares the plots of functions $A({\tilde H}_0)$ for simple and arc bending.
     It is evident that formation of the arc entails a greater accumulation of elastic energy, and therefore, the field under which an arc configuration first becomes possible is higher.
     The conclusion about absolute stability of the linear configuration of an ideal cantilever with respect to infinitely small perturbations is also valid in this case, especially since the arc deformation is more energy-consuming than a simple bending.

\subsection{Two-parameter approximation} \label{sec:05.2}
     The trial configuration function for the arc bending is also chosen in the form of a fourth-degree polynomial:
\begin{equation}  \label{eq:15}
    \tilde{u}_z={\tilde x}^2(1-{\tilde x})(B{\tilde x}+C),
\end{equation} 
under condition that only upward deflection is allowed.
     Evidently, the requirement for the free end not to detach from the basic ($z=0$) plane entails reduction of the number of variational coefficients from three to two.

     As already noted, the material and geometric parameters of a real MAE rod as well as the uniformity of the field cannot be perfect, and one would not encounter absolute stability of cantilevers in plain experiments.
     Because of that, the arc deformation seems to occur, as if spontaneously, when the field grows up to a certain value that, in fact, is determined by individual characteristics of the given cantilever and the field source.
     Since our approach cannot account for such situations, here, as in section \ref{sec:04}, we limit ourselves to a test calculation, which involves deliberate introduction of a controlled perturbation.
     The chosen kind of perturbation is an upward lifting of the middle point of the cantilever by $\delta=(\Delta z/L)_{{\tilde x}=1/2}$.
     Curve \emph{2} in Fig.\ \ref{fig:fig_07} presents the dependence ${\tilde H}_{0\ast}(\delta)$ resulting from 'provocation' of the cantilever into arc bending; the data is obtained from numerical calculations with two-parameter variational function (\ref{eq:15}).
      Evidently, this also confirms the conclusion drawn from the qualitative analysis that, compared to simple bending, arc bending requires a greater increment of energy for its occurrence, i.e., a higher applied field: curves \emph{2} goes above curve \emph{1} in Fig.\ \ref{fig:fig_07}.

\section{Discussion} \label{sec:06}
     The presented theoretical analysis shows that a perfect magnetically soft MAE cantilever that resides under gravity field on a solid horizontal plane is absolutely stable against infinitesimal perturbations whatever the magnetic field applied vertically.
     Meanwhile, as soon as the field becomes sufficiently strong, there emerges an alternative curvilinear configuration, which, provided the cantilever assumes it, grants a substantial gain in magnetostatic energy and, thus, one more stable state.
     The two configurations are separated by a potential barrier that any field, whatever strong, cannot eliminate, less so it is possible due to thermal fluctuations: the system is too massive.
     In the model (ideal) situation this means that only an externally exerted mechanical perturbation may move the cantilever from one state to the another one.
     
     In real situation the role of perturbation might (and would) be played by ubiquitous imperfections of the quality of the MAE, those of the rod shape and the non-uniformities of the field.
     The task of estimating the energy of a perturbation necessary for inducing the above-mentioned configuration change transition expands, in fact, into a great number of particular magnetomechanical problems each of which should consider an imperfectness of a precise type.
     This scope is beyond the frame of the present work.
     However, irrespective of the origin of perturbation, one may obtain the sought for estimate from considering the height of the energy barrier that separates the two above-mentioned minima.
     Evidently, this barrier emerges when the field magnitude attains some reference value where the alternative minimum turns up yet as an inflection fold at the ${\tilde U}(A,B,C)$ potential surface.
     At that moment, the barrier hight is yet formally infinite but it begins to reduce as the field grows, and the fold in the $(A,B,C)$ space transforms into a ridge that surrounds the location occupied by the minimum with $A=B=C=0$, that formerly was the only one.
     
     As an example, let us estimate the barrier height for the three-parameter model in the case of simple bending under assumptions ${\tilde u}_z\ll1$.
     For this purpose, the fourth-order term in (\ref{eq:06}) may be neglected, so that the accordingly truncated expression is
\begin{equation} \label{eq:16}
   \tilde{U^\prime}=\int_0^1d{\tilde x}\left\{{\textstyle \frac12}\beta\left(\frac{d^{2}{\tilde u}_z}{d{\tilde x}^{2}}\right)^{\!2}\!\!-{\tilde H}_0^2\left(\frac{d{\tilde u}_z}{d{\tilde x}}\right)^{\!\!2}+\alpha{\tilde u}_z\right\}.
\end{equation}
     Substituting there profile (\ref{eq:10}) one finds after integration a function $U^\prime(A,B,C)$.
     Taking its derivatives with respect to the variational parameters and setting the result to zero, gives a set of linear equations:
\begin{eqnarray}
   \frac{\partial{\tilde U}^\prime}{\partial A}=-{\tilde H}_0^2\left({\textstyle \frac{32}{7}}a-4b-{\textstyle \frac{16}{5}}c\right) + {\textstyle \frac{144}{5}}a\beta+{\textstyle \frac{1}{5}}\alpha+18b\beta+8\beta c=0, \nonumber \\[4pt]
   \frac{\partial{\tilde U}^\prime}{\partial B}=-{\tilde H}_0^2\left(4a-{\textstyle \frac{18}{5}}b-3c\right)+18a\beta+{\textstyle \frac{1}{4}}\alpha+12b\beta+6\beta c=0,
                                                                \label{eq:17} \\[4pt]
    \frac{\partial{\tilde U}^\prime}{\partial C}=-{{\tilde H}_0^2\left({\textstyle \frac{16}{5}}a-3b-{\textstyle \frac{8}{3}}c\right)+8a\beta+{\textstyle \frac{1}{3}}\alpha+6b\beta+4 \beta c=0, \nonumber }
\end{eqnarray}     
to find the extremum of ${\tilde u}_z$.
     The solution of the set (\ref{eq:17}) may be written down analytically but the corresponding expression is rather lengthy, and because of that we go directly to the result of its substitution in (\ref{eq:16}) that is
\begin{equation}  \label{eq:18}
    \tilde U_{\ast}=\frac{\alpha^{2}\cdot\left(13 {\tilde H}_0^{4}-756{\tilde H}_0^{2} \beta+3780\beta^{2}\right)}{96\cdot\left(2{\tilde H}_0^{6}-135{\tilde H}_0^{4}\beta+1440{\tilde H}_0^{2} \beta^{2}-1575 \beta^{3}\right)}.
\end{equation}     
     Note that this expression (\emph{i}) is obtained from the truncated form of energy (\ref{eq:16}) and (\emph{ii}) the basic requirement ${\tilde u}_z\geq0$, i.e., only upward displacements are possible, is removed here.
     Because of that, the denominator of formula (\ref{eq:18}) predicts non-physical results within the field range $[{\tilde H}_s,{\tilde H}_g]$ where ${\tilde H}_s$ and ${\tilde H}_g$ are, respectively, the smallest and the greatest real roots of the denominator polynomial; inside this interval the denominator has several poles as a function of ${\tilde H}_0^2$ and, thus, changes sign.
     
     However, expression (\ref{eq:18}) is not entirely useless.
     There are two limits---${\tilde H}_0<{\tilde H}_s$ and ${\tilde H}_0>{\tilde H}_g$---where it has a definite sign. 
     The first limit, although formally correct, yields negative ${\tilde u}_z$, i.e., downward deflexions, and corresponds to an energy minimum.
     Evidently, this prediction is not relevant for the cantilever under horizontal-plane constraint and has emerged just because expression (\ref{eq:16}) allows for the displacements of any sign.
     Meanwhile the second limit (${\tilde H}\rightarrow\infty$):
 \begin{equation}  \label{eq:19}
   {\tilde U}_{\ast}=\frac{13}{192}\frac{\alpha^2}{{\tilde H}^2}+O\left({\tilde H}_0^{-4}\right),
\end{equation}    
adds to the analysis of the problem.
     It corresponds to the energy maximum and thus establishes that the height of potential barrier that surrounds the initial point $A=B=C=0$ (horizontal configuration), at strong fields decreases proportionally to ${\tilde H}_0^{-2}$, i.e., as it had been surmised above, the barrier tends to zero just asymptotically.
     
     Putting apart the details, let us consider the problem of a cantilever under gravity in a more general context.    
     When the MAE cantilever is single, it may be considered as an element of a magnetically controlled manipulator, gripper, or valve.
     On the other hand, a collection of miniature horizontal cantilevers---referred to as cilia or brushes---can form a coating that is able to move the objects across it, create or suppress flows in adjacent layers of fluid, or serve for filtering suspensions.
     Most studies aimed at applications, consider the coatings where, in the initial state, the cilia are oriented perpendicular to the substrate surface \cite{ZhWa_SAB_18,ZhGu_AIS_21}; a typical mode of their use is counteracting normal pressure (load) on such a coating by enhancing the applied field.
     Horizontal (tangential) cilia, unlike vertical (normal) ones, in the initial state but slightly distort the smoothness of the surface—they `rear up' only in response to the applied field.
     In this sense, the response of such a coating is opposite to that of a coating with normal cilia, and the same applies to the sets of MAE mini-rods (brushes) used as filters in flows with impurities.
     The results of our modelling infer that with a moderately strong field, the tangential cilia can be erected almost vertically to the surface.
     Then the roughness of the coating can be controllably reduced, thereby decreasing, for example, the near-wall dissipation of energy in the fluid flowing through the channel.

     Finally, we would like to emphasise that the main goal of the presented results is to provide an understanding of the observed magneto-deformation effect and reveal the non-trivial nature of the transition underlying it.
     On the one hand, a horizontal MAE cantilever, in principle, is absolutely stable with respect to the applied field due to the `energy pocket' caused by gravity and the confining surface.
     On the other hand, the always-present imperfections in the geometry and material parameters of the rod and the field source inevitably cause a breakdown of this stability.
     For this reason, there is no general expression for the `erection' field ${\tilde H}_{0*}$ of the cantilever in the considered situation; this value turns out to be an individual characteristic in each case.
     However, it is evident that the transition field, for instance, in an ensemble of cilia, can always be synchronised if a given distortion is introduced into the configuration of each one, thereby setting a uniform ${\tilde H}_{0*}$ for all.
     Moreover, the value of ${\tilde H}_{0*}$ could be pre-programmed while manufacturing with the aid of calibrating curves like those shown in Fig.\ \ref{fig:fig_07} or their analogs, if more accurate calculations would be available.

\section*{Acknowledgements}
This work was carried out within the framework of project AAAA-A20-120020690030-5.

\appendix

\section*{Appendix. Magnetic energy of a magnetically soft spheroid}
     Consider an ellipsoid of revolution with semi-axes $q_1$ (major) and $q_2$ (minor, the radius of the cross-section) made of magnetically soft isotropic material with magnetic susceptibility $\chi$.
     Under a uniform external field $\bm{H}_0$, the magnetic field inside the ellipsoid is also uniform and is given by
\begin{equation} \label{eq:A01}
   \bm{H}=\bm{H}_0-4\pi{\mathbf{N}}\!\cdot\!\bm{M},
\end{equation}
where \(\bm{M}\) is the magnetization, and \({\mathbf{N}}=2N_{nn}\bm{n}\bm{n}+N_{\tau\tau}\bm{\tau}\bm{\tau}\) is the demagnetizing tensor expressed in terms of the unit vectors along the principal axes.
     The components of this diagonal tensor depend on the aspect ratio $q_1/q_2$ and are normalized by the relation $2N_{nn} + N_{\tau\tau}=1$, where $\bm{\tau}$ is the direction along the principal axis, and $\bm{n}$ is perpendicular to it.

     Assuming a linear magnetization law $\bm{M}=\chi\bm{H}$ in (\ref{eq:A01}) and solving the resulting equation for $\bm{H}$, one finds
\begin{equation}  \label{eq:A02}
   \bm{H}=(\mathbf{I}+4\pi\chi\mathbf{N})^{-1}\!\cdot\!\bm{H}_0,
\end{equation}
where
\begin{equation}
   (\mathbf{I}+4\pi\chi\mathbf{N})^{-1} = \frac{2\bm{n}\bm{n}}{1+4\pi\chi N_{nn}} + \frac{\bm{\tau}\bm{\tau}}{1+4\pi\chi N_{\tau\tau}};
\end{equation}
here $\mathbf{I}$ is the identity tensor.

     The magnetic energy, given by general formula $U_{\rm mag}=-\frac{1}{2} \bm{M}\!\cdot\!\bm{H}_0 V_e$, where $V_e$ is the volume of the ellipsoid, in this situation takes the form
\begin{equation}  \label{eq:A03}
   U_{\rm mag}=-{\textstyle \frac{1}{2}}\chi \bm{H}_0(\mathbf{I}+4\pi\chi\mathbf{N})^{-1}\bm{H}_0 V_e=
-{\textstyle \frac{1}{2}}\chi\!\left[\frac{(\bm{n}\!\cdot\!\bm{H}_0)^2}{1\!+\!4\pi\chi N_{nn}}+\frac{(\bm{\tau}\!\cdot\!\bm{H}_0)^2}{1\!+\!4\pi\chi N_{\tau\tau}}\right]\!\!V_e.
\end{equation}
     If the vectors $\bm{n}$ and $\bm{H}_0$ form an angle $\theta$, then the angle between $\bm{\tau}$ and $\bm{H}_0$ is $\pi/2-\theta$; in these notations one has
\begin{equation}  \label{eq:A04}
   U_{\rm mag}=-{\textstyle \frac{1}{2}}\chi H_0^2 V_e \frac{(1+4\pi\chi N_{\tau\tau})+4\pi\chi \sin^2\theta (N_{nn}-N_{\tau\tau})}{(1+4\pi\chi N_{\tau\tau})(1+4\pi\chi N_{nn})}.
\end{equation}

     Upon retaining only the angle-dependent part of the energy—since this is the term that changes when the rod bends—the volume density writes as
\begin{eqnarray}   \label{eq:A05}
   W_{\rm mag}=-2\pi\chi^2 H_0^2 \frac{\sin^2\theta (N_{nn}-N_{\tau\tau})}{(1+4\pi\chi N_{\tau\tau})(1+4\pi\chi N_{nn})}=
   \\ [6pt] \hspace{45mm} -S(\mathbf{N},\chi) H_0^2 \sin^2\theta=-S(\mathbf{N},\chi) \left(\bm{\tau}\!\cdot\!\bm{H}_0\right)^2, \nonumber
\end{eqnarray}
thus defining the form factor of the ellipsoid in the form
\begin{equation}  \label{eq:A06}
   Q(\mathbf{N},\chi)=\frac{2\pi \chi^2 (N_{nn}-N_{\tau\tau})}{(1+4\pi\chi N_{\tau\tau})(1+4\pi\chi N_{nn})}.
\end{equation}  \label{eq:A07}
     In the special case of a very long rod, assuming $N_{\tau\tau} = 0\) and \(N_{nn} = 0.5$, expression (\ref{eq:A06}) reduces to
\begin{equation} \label{eq:A08}
   Q = \pi\chi^2/(1+2\pi\chi).
\end{equation}
     It should be noted that the above-presented derivation of the approximate expression (\ref{eq:A08}) conceptually repeats the result obtained by A.\ Cebers in the calculation of the forces acting on an element of length of an elongated droplet of magnetic fluid \cite{Cebe_PRE_02} or a filament formed by a chain of interconnected magnetically soft particles \cite{Cebe_JP-CM_03}.

\bibliographystyle{unsrt}
\bibliography{erection_bib_v1}
\end{document}